\documentclass[letterpaper, 10 pt, conference]{ieeeconf}  
\IEEEoverridecommandlockouts                              
\usepackage{amsmath,amssymb,amsthm,mathtools,bm}
\usepackage{graphicx,booktabs,multirow,array}
\usepackage[hidelinks,bookmarks=false,draft]{hyperref}   
\usepackage[capitalize,noabbrev]{cleveref}
\usepackage{microtype}
\usepackage{needspace}
\graphicspath{{figs/}}
\newcommand{\NumEoneGrid}{0,1/32,1/16,1/8,1/4,1/2,1}

\newcommand{\NumEoneSlopes}{1.05 to 1.08}
\newcommand{\NumOptTermP}{4.3\times10^{6}}

\newcommand{\NumTailFiveTwelve}{7.78\times10^{-9}}

\newcommand{\NumEoneMeasured}{$0.458\pm0.037$}
\newcommand{\NumEoneUniform}{$3.19\times10^{6}$}
\newcommand{\NumEoneOccupancy}{$5.69\times10^{5}$}
\newcommand{\NumEoneEnvelope}{$1.47\times10^{5}$}
\newcommand{\NumOccGain}{5.6}
\newcommand{\NumEoneWeightedPct}{19\%}
\newcommand{\NumLamCross}{0.2147}

\newcommand{\NumObsNominal}{$0.47$ and $1.2$}
\newcommand{\NumObsLearned}{$0.18$ and $1.3$}
\newcommand{\NumObsDetuned}{$0.0081$ and $3.4$}

\newcommand{\NumObsErrRange}{$0.051$ to $0.165$}

\newcommand{\NumAlphaCert}{2.5\times10^{-4}}
\newcommand{\NumInitRetD}{19.23}
\newcommand{\NumChgND}{$-2.83$ $[-3.59,-2.15]$}
\newcommand{\NumChgLD}{$-0.20$ $[-0.35,-0.07]$}
\newcommand{\NumChgXD}{$-0.10$ $[-0.21,+0.00]$}
\newcommand{\NumSimRiseND}{$+3.03$}
\newcommand{\NumInitRetW}{17.64}
\newcommand{\NumChgNW}{$-4.71$ $[-5.27,-4.11]$}
\newcommand{\NumChgLW}{$+1.40$ $[+1.19,+1.60]$}
\newcommand{\NumChgXW}{$+1.31$ $[+1.08,+1.53]$}
\newcommand{\NumSimRiseNW}{$+3.15$}

\newcommand{\NumRMSEqInitD}{0.052}

\newcommand{\NumEffRetD}{$+0.63$ $[+0.28,+1.05]$}
\newcommand{\NumEffRetDt}{$[+0.16,+1.10]$}

\newcommand{\NumLQRDet}{19.60}

\newcommand{\NumLQRGain}{33.7, 11.0}

\newcommand{\NumEffRetW}{$+4.48$ $[+3.89,+5.05]$}
\newcommand{\NumEffRetWt}{$[+3.80,+5.17]$}

\newcommand{\NumDNormMNR}{$15$ $[3.3,29]$}
\newcommand{\NumDInnerNNR}{$-23$ $[-53,-4.8]$}
\newcommand{\NumDInnerLNR}{$14$ $[3.4,26]$}

\newcommand{\NumDNormMLR}{$7.8$ $[1.2,17]$}

\newcommand{\NumDNormMNRW}{$96$ $[85,110]$}
\newcommand{\NumDInnerNNRW}{$-37$ $[-52,-19]$}
\newcommand{\NumDInnerLNRW}{$100$ $[92,110]$}

\newcommand{\NumDNormMLRW}{$17$ $[8.1,26]$}

\newcommand{\NumDInnerNegNNR}{12}

\newcommand{\NumDDatasets}{12}

\newcommand{\NumDInnerPosLNR}{12}

\newcommand{\NumDMargNegNNRW}{12}

\newcommand{\NumDMargNegResNNRW}{11}

\newcommand{\NumDInnerNegNNRW}{10}

\newcommand{\NumDInnerNegResNNRW}{7}
\newcommand{\NumDMargPosLNRW}{12}

\newcommand{\NumDMargPosResLNRW}{12}

\newcommand{\NumDInnerPosLNRW}{12}

\newcommand{\NumDInnerPosResLNRW}{12}

\newcommand{\NumDNormMDiffNR}{$7.4$ $[0.3,17]$}
\newcommand{\NumDNormMDiffNRW}{$80$ $[65,96]$}

\newcommand{\NumPoneKappaH}{8.6}
\newcommand{\NumPoneKappaTot}{22.5}
\newcommand{\NumPoneUniformN}{3.31\times10^{7}}
\newcommand{\NumKappaTrivial}{0.316}

\newcommand{\NumDeployKappaTwo}{0.05}

\newcommand{\NumPoneKappaLTot}{0.9}
\newcommand{\NumPoneKappaLRange}{0.5 to 1.6}
\newcommand{\NumPoneUniformL}{1.3\times10^{6}}
\newcommand{\NumPoneModelTermN}{6.6\times10^{7}}
\newcommand{\NumPoneModelTermL}{2.7\times10^{6}}
\newcommand{\NumEfourKs}{16, 32, 64, 128, 256}
\newcommand{\NumEfourPols}{3}
\newcommand{\NumEfourCells}{15}
\newcommand{\NumEfourMinFactor}{200}

\newcommand{\NumNomInitRetD}{11.19}
\newcommand{\NumNomRMSEqD}{0.22}

\newcommand{\NumNomChgND}{$+0.81$ $[+0.44,+1.18]$}

\newcommand{\NumNomChgXD}{$+3.48$ $[+2.85,+4.10]$}

\newcommand{\NumNomLQRD}{12.84}

\newcommand{\NumNomRunsD}{12}

\newcommand{\NumNoiseScales}{1/16, 1/4, 1, 4}

\newcommand{\NumNoiseZRange}{$0.41$ to $0.55$}

\newcommand{\NumNoiseKappaRange}{$0.54$ to $34.7$}

\newcommand{\NumNoiseBoundFactor}{64}

\newcommand{\NumFDMaxZ}{0.8}
\newcommand{\NumFDSteps}{$0.02$, $0.05$, $0.1$}

\hypersetup{pdftitle={Policy-Gradient Transfer under a Shared State Estimator},pdfauthor={Dongming Wang and Wei Ren},
  pdfsubject={Policy-gradient transfer for partially observed control},pdfkeywords={policy gradient, model-based reinforcement learning, output-feedback control, model uncertainty}}

\theoremstyle{plain}
\newtheorem{theorem}{Theorem}
\newtheorem{lemma}{Lemma}
\newtheorem{proposition}{Proposition}
\newtheorem{corollary}{Corollary}
\theoremstyle{definition}
\newtheorem{assumption}{Assumption}

\crefname{assumption}{Assumption}{Assumptions}
\Crefname{assumption}{Assumption}{Assumptions}

\newcommand{\R}{\mathbb{R}}
\newcommand{\E}{\mathbb{E}}
\newcommand{\N}{\mathbb{N}}
\newcommand{\U}{\mathcal{U}}
\newcommand{\M}{\mathcal{M}}
\newcommand{\Dx}{D_{x}}
\newcommand{\real}{M}
\newcommand{\sm}{\widehat{M}}
\newcommand{\xhat}{\widehat{x}}
\newcommand{\cem}{\dagger}
\newcommand{\Rmax}{R_{\max}}
\newcommand{\mub}{\bar\mu}
\newcommand{\KL}{\mathrm{KL}}

\newcommand{\Spd}{\mathbb{S}^{n_x}_{++}}
\DeclareMathOperator{\diag}{diag}

\begin{document}

\title{\LARGE \bf Policy-Gradient Transfer under a Shared State Estimator}

\author{Dongming Wang$^{1}$ and Wei Ren$^{1}$
\thanks{$^{1}$Dongming Wang and Wei Ren are with the Department of Electrical and Computer Engineering, University of California, Riverside, CA 92521, USA
        {\tt\small wdong025@ucr.edu, ren@ece.ucr.edu}}%
}

\maketitle
\thispagestyle{empty}
\pagestyle{empty}

\begin{abstract}
This paper considers the problem of policy-gradient transfer from a simulator to a physical plant when both are controlled through the same fixed state estimator. The plant and the simulator are partially observed nonlinear systems with additive Gaussian noise of common known covariance whose dynamics and observation models differ, and the controller is a squashed Gaussian policy with uniformly bounded latent mean and first two parameter derivatives. We derive a bound on the difference between the plant and simulator policy gradients that is linear in the noise-weighted model error and requires no estimator-accuracy guarantee, because the shared estimator is a common post-processing of the measurements and the data-processing inequality compares the trajectory laws without an estimation-error term. Based on this bound, we derive an occupancy-weighted refinement, a sufficient condition for the simulator gradient to be an ascent direction for the plant, and a bound on the expected squared plant gradient at a random iterate of a finite simulator-training run with optimization, truncation and model-error terms. Simulation results on a scalar system and a pendulum are presented to validate the theoretical results.
\end{abstract}

\section{Introduction}\label{sec:intro}

Learning-based controllers are commonly trained in a simulator and then deployed on hardware, where the controller acts on the output of an onboard state estimator~\cite{kaufmann2023}. When a policy-gradient method~\cite{williams1992,baxter2001} estimates the gradient in a simulator, a sufficiently small positive step improves the plant objective if the simulator and plant policy gradients at the same parameter have a positive inner product. This condition holds when the distance between the two gradients is smaller than the simulator gradient norm, and model mismatch can violate it. Consequently, a policy update that improves the simulator return may decrease the plant return. We study how errors in the simulator's dynamics and observation models affect the discrepancy between the plant and simulator policy gradients.

Return bounds under model error~\cite{asadi2018,janner2019} and the sim-to-real performance guarantee for linear-quadratic-Gaussian systems in~\cite{hu2023sim2real} do not directly control the policy-gradient discrepancy studied here. For fully observed systems, the stationarity reached with policy gradients from an inaccurate deterministic model grounded by real trials is bounded by the model-derivative error in~\cite{abbeel2006}, on-policy model errors that bias model-based policy improvement are corrected with real data in~\cite{frohlich2022}, and the bias of reparameterized model-based gradients is bounded by model and critic errors in~\cite{zhang2023rp}. In~\cite{abachi2020}, a policy-aware model loss is derived from a gradient approximation that pairs the model's state occupancy with the true action values, and in~\cite{doro2020} the model is learned with a relative-entropy loss weighted by the policy's occupancy and score magnitude, with the gradient error bounded by pairing the true occupancy with model action values. Under partial observation, approximate information states~\cite{subramanian2022}, actor-critic methods with finite memory~\cite{cayci2024} and jointly learned internal states~\cite{kar2026} analyze learning within one environment, filter-kernel perturbations and the performance of policies transferred between perturbed models are bounded in~\cite{demirci2025filter}, and learning with an estimator in the loop is demonstrated empirically in~\cite{wang2024ajlatt}. To our knowledge, no existing result bounds the difference between the exact closed-loop policy gradients of a plant and a simulator that share a fixed state estimator in terms of forward dynamics and observation-model errors, without model derivatives or an estimator-accuracy assumption.

Model mismatch can alter the measurements and, through the estimator's memory, affect subsequent control actions. Consequently, using the same estimator in the plant and simulator does not ensure identical estimator trajectories. Comparing the resulting policy gradients is further complicated by termination at operating limits, which can make the realized return a discontinuous functional of the trajectory, and by the unbounded Gaussian policy score. We address these difficulties using likelihood-ratio identities for the policy gradient and second-moment bounds for the score. This approach accommodates a general fixed measurable estimator without requiring an estimation-error bound.

The main purpose of this paper is to bound the gradient discrepancy with no assumption on the fixed shared estimator beyond measurability. First, we show that the difference between the exact plant and simulator gradients is at most a constant multiple of the model error, the largest noise-weighted discrepancy between the dynamics maps and between the observation maps, where the constant depends only on the reward range, the discount and the score moments of the policy. The key observation is that the same Borel map processes the measurements in both environments, so the data-processing inequality bounds the divergence of the trajectory laws without any additional regularity or accuracy assumption on that map. Second, we show that the worst-case model error can be replaced by errors weighted by the plant's discounted occupancy, which relates gradient transfer to model accuracy along visited trajectories and complements the model-learning weights of~\cite{doro2020}, and that the simulator gradient is an ascent direction for the plant if its distance from the plant gradient is smaller than its norm. Third, we show that a likelihood-ratio method that does not interact with the plant after identification yields a bounded expected squared \emph{plant} gradient, with separate optimization, truncation and model-discrepancy terms.

\section{Problem Formulation}\label{sec:problem}
\subsection{Two environments, one observer}
We compare the real plant $\real$ and the simulator $\sm$ at the same policy parameter. Their dynamics and observation maps may differ, while the policy, state estimator (observer), reward, stopping rule, initial law and noise covariances are common to both environments. For $E\in\{\real,\sm\}$ the state and observation satisfy
\begin{equation}\label{eq:env}
x_{t+1}=f_E(x_t,u_t)+w_t,\quad y_{t+1}=h_E(x_{t+1})+v_{t+1},
\end{equation}
with $x_t\in\R^{n_x}$, $u_t\in\U\subset\R^{n_u}$, $y_t\in\R^{n_y}$, independent Gaussian noises $w_t\sim\mathcal N(0,Q)$, $v_t\sim\mathcal N(0,R_v)$, and $Q\succ0$, $R_v\succ0$ known and common to both environments. Vector norms are Euclidean, matrix norms induced, $\|a\|_W^2=a^\top Wa$, and relative entropy uses natural logarithms.

\begin{assumption}[Environments]\label{ass:env}
$f_E$ and $h_E$ are Borel measurable. For a Borel physical gate $\Dx\subseteq\R^{n_x}$ the noise-weighted local model errors
\begin{equation}\label{eq:eps}
\kappa_f^2:=\sup_{\Dx\times\U}\|f_{\real}-f_{\sm}\|_{Q^{-1}}^2, \kappa_h^2:=\sup_{\Dx}\|h_{\real}-h_{\sm}\|_{R_v^{-1}}^2
\end{equation}
are finite, and $\kappa^2:=\kappa_f^2+\kappa_h^2$.
\end{assumption}
The estimator is a fixed Borel map $F:\M\times\U\times\R^{n_y}\to\M$ on a standard Borel memory space $\M$ with a Borel readout $\chi:\M\to\R^{n_x}$; both maps are independent of $\theta$, the same $F$ runs in both environments, and the actor receives only $\chi(m_t)$. For the extended Kalman filter (EKF) used in the simulations, the memory is $m=(\xhat,P)$ in $\M=\R^{n_x}\times\Spd$, and the readout is $\chi(m)=\xhat$. The map $F$ is the EKF recursion with a Joseph-form covariance update, constructed from fixed dynamics and observation maps $(f_F,h_F)$ that are $C^1$ and finite everywhere. With $P_0\succ0$ and $Q,R_v\succ0$, the innovation covariance is invertible and $P_t\succ0$.

The analysis uses only measurability of $F$, which may be built from a model other than the training simulator; the augmented state $Z_t=(x_t,m_t)$ is Markov, and the same stopping rule acts on it in both environments.

\subsection{Stopping, reward, objective}\label{sec:stop}
Fix a Borel live set $D\subseteq\Dx\times\M$, a cemetery state $\cem\notin\Dx\times\M$, and a common initial law $\nu_0$ of the pair $(x_0,m_0)$ on $D$; no observation is drawn at time $0$. From a live state $Z_t=(x_t,m_t)$ and action $u_t$, draw $x_{t+1}\sim\mathcal N(f_E(x_t,u_t),Q)$. If $x_{t+1}\notin\Dx$, set $Z_{t+1}=\cem$. Otherwise draw $y_{t+1}\sim\mathcal N(h_E(x_{t+1}),R_v)$, set $m_{t+1}=F(m_t,u_t,y_{t+1})$, and let $Z_{t+1}=(x_{t+1},m_{t+1})$ when this pair belongs to $D$ and $Z_{t+1}=\cem$ otherwise. The cemetery is absorbing. Denote the stopped kernel by $\mathcal K_E$ and the exit time by $\tau=\inf\{t:Z_t=\cem\}$. A Borel reward $r:\R^{n_x}\times\U\to[0,\Rmax]$ is collected at live steps, $r_t=r(x_t,u_t)\mathbf 1\{t<\tau\}$, so rewards and scores vanish at the cemetery; functions of a live state are extended by zero to $\cem$. For a discount $\gamma\in(0,1)$ the objective and its $H$-step truncation are
\begin{equation}\label{eq:J}
J_E(\theta)=\E_E\sum_{t\ge0}\gamma^tr_t,\qquad J_E^H(\theta)=\E_E\sum_{t<H}\gamma^tr_t,
\end{equation}
with $J_E^\infty=J_E$ and empty sums equal to zero. Because the physical gate is tested before an observation is requested, the observation-model error enters only at physical states inside $\Dx$. A live set $D$ that depends on the true state is an instrumented boundary. Conditional on the identification data, the environment and estimator maps, the gates, the initial law, the reward, the discount and the noise covariances are fixed and independent of $\theta$, and the process-noise, measurement-noise and actor-noise draws are mutually independent over time and of $Z_0$; if identification establishes the bounds $(\kappa_f,\kappa_h)$ on a high-probability event, the conclusions apply on that event.

\subsection{Actor}
\begin{assumption}[Actor]\label{ass:actor}
Actions are $u=\psi(z)$ with $\psi(z)=u_{\max}\odot\tanh(z)$, $u_{\max,j}>0$, and $\U=\prod_j(-u_{\max,j},u_{\max,j})$. The latent is $z=\mu_\theta(\chi(m))+\Sigma^{1/2}\xi$ with $\xi\sim\mathcal N(0,I)$ and $\Sigma=\diag(\sigma_1^2,\dots,\sigma_{n_u}^2)\succ0$ fixed. For every $\theta$ the map $s\mapsto\mu_\theta(s)$ is Borel; for every $s$, $\theta\mapsto\mu_\theta(s)$ is $C^2$ on $\R^p$; and there are constants $\mub_j,B_{1,j},B_{2,j}$ with $|\mu_{\theta,j}(s)|\le\mub_j$, $\|\nabla_\theta\mu_{\theta,j}(s)\|\le B_{1,j}$, $\|\nabla^2_\theta\mu_{\theta,j}(s)\|\le B_{2,j}$ for all $\theta,s,j$. At $\cem$ the latent is drawn from a fixed $\theta$-independent law and the score is zero.
\end{assumption}
These bounds give the score-moment control of~\cite{yuan2022} and an integrable envelope for differentiating the expected stopped reward; continuity in $\theta$ makes $(\theta,s)\mapsto\mu_\theta(s)$ jointly measurable. At a live step the score is
\begin{equation}\label{eq:score}
g_t=\nabla_\theta\log \pi_\theta(z_t\mid s_t)=\nabla_\theta\mu_\theta(s_t)^\top\Sigma^{-1}\bigl(z_t-\mu_\theta(s_t)\bigr),
\end{equation}
where $\nabla_\theta\mu_\theta(s)$ is the $n_u\times p$ Jacobian and $\nabla_\theta\mu_{\theta,j}(s)$ its $j$th row as a $p$-vector, with $s_t=\chi(m_t)$, and $g_t=0$ after absorption. Since the squashing map is a bijection onto the open action box with a $\theta$-independent Jacobian, \eqref{eq:score} is also the score of the action density. Let $S_t=\sum_{j\le t}g_j$. At live steps define $\Upsilon_t=\nabla_\theta^2\log \pi_\theta(z_t\mid s_t)$, and set $\Upsilon_t=0$ after absorption. Let $\mathcal F_t$ be the $\sigma$-field generated by $(Z_0,z_0,\dots,Z_t)$. The constants $G^2:=\sum_jB_{1,j}^2/\sigma_j^2$ and $L_g:=\sum_j(B_{1,j}^2/\sigma_j^2+\sqrt{2/\pi}\,B_{2,j}/\sigma_j)$ bound the conditional score moments.

\section{Policy-Gradient Transfer}\label{sec:theory}
In this section, we bound the difference between the plant and simulator policy gradients and derive its consequences.
\subsection{Differentiating the stopped objective}\label{sec:diff}
\begin{proposition}[Actor constants]\label{prop:actor}
Under \Cref{ass:actor}, for every $\theta$ and live $s$, $\E_{z\sim\mathcal N(\mu_\theta(s),\Sigma)}\|\nabla_\theta\log \pi_\theta(z\mid s)\|^2\le G^2$ and $\E\|\nabla_\theta^2\log \pi_\theta(z\mid s)\|\le L_g$. For the scalar actor $\mu_\theta(s)=\mub\tanh(\theta^\top\phi(s))$ with $\|\phi\|\le1$, one may take $B_1=\mub$ and $B_2=4\mub/(3\sqrt3)$.
\end{proposition}
\begin{proof}
Write $a_j=\nabla_\theta\mu_{\theta,j}(s)/\sigma_j$ and $\xi_j=(z_j-\mu_{\theta,j}(s))/\sigma_j$, so that $g=\sum_ja_j\xi_j$ by \eqref{eq:score}. Since the coordinates of $\xi$ are independent standard Gaussians, $\E\|g\|^2=\sum_j\|\nabla_\theta\mu_{\theta,j}\|^2/\sigma_j^2\le G^2$. Differentiating \eqref{eq:score} gives $\nabla^2_\theta\log \pi_\theta=-\sum_j\nabla\mu_j\nabla\mu_j^\top/\sigma_j^2+\sum_j(z_j-\mu_j)\nabla^2\mu_j/\sigma_j^2$, and $\E|z_j-\mu_j|=\sigma_j\sqrt{2/\pi}$ gives $L_g$. For $\mu=\mub\tanh(\zeta)$ with $\zeta=\theta^\top\phi$, $\nabla_\theta\mu=\mub\operatorname{sech}^2(\zeta)\phi$ and $\nabla^2_\theta\mu=-2\mub\tanh(\zeta)\operatorname{sech}^2(\zeta)\phi\phi^\top$ with $\sup_{v\in[0,1]}(v-v^3)=2/(3\sqrt3)$ give $B_1$ and $B_2$.
\end{proof}

\begin{lemma}[Score identities and smoothness]\label{lem:score}
Under \Cref{ass:env,ass:actor}, for every $E$, $t$ and $\theta$, the maps $\theta\mapsto\E_E[r_t]$ are $C^2$ with
\begin{gather}
\nabla_\theta\E_E[r_t]=\E_E[r_tS_t],\nonumber\\ \nabla_\theta^2\E_E[r_t]=\E_E\Bigl[r_t\Bigl(S_tS_t^\top+\textstyle\sum_{j\le t}\Upsilon_j\Bigr)\Bigr],\label{eq:lr}
\end{gather}
$\E_E[g_j\mid\mathcal F_j]=0$ and $\E_E\|S_t\|^2\le(t+1)G^2$. Moreover $J_E$ is twice continuously differentiable with $\nabla J_E=\sum_t\gamma^t\nabla\E_E[r_t]$ and $\|\nabla^2J_E(\theta)\|\le L_J:=\Rmax(G^2+L_g)/(2(1-\gamma)^2)$ for all $\theta$, so $\nabla J_E$ is $L_J$-Lipschitz.
\end{lemma}
\begin{proof}
Fix $E$ and $\theta_0$, and let $\mathbb P^E_{\theta_0,t}$ be the law of the history $\omega_t=(Z_0,z_0,\dots,Z_t,z_t)$ under $\theta_0$. Only the latent laws depend on $\theta$; they are nondegenerate Gaussians at live steps and common at $\cem$, so the history law under $\theta$ has the finite likelihood ratio
\[
\Lambda_t(\theta)=\prod_{j\le t:\,Z_j\ne\cem}\frac{\pi_\theta(z_j\mid s_j)}{\pi_{\theta_0}(z_j\mid s_j)},\ \ \pi_\theta(z\mid s)=\varphi_\Sigma(z-\mu_\theta(s)),
\]
with respect to $\mathbb P^E_{\theta_0,t}$, and $\E_E[r_t]=\E_{\theta_0}[\Lambda_t(\theta)r_t]$. For a live step let $\varsigma_j=\Sigma^{-1/2}(\mu_\theta(s_j)-\mu_{\theta_0}(s_j))$ and $\xi_j=\Sigma^{-1/2}(z_j-\mu_{\theta_0}(s_j))$. Then $\log(\pi_\theta/\pi_{\theta_0})(z_j\mid s_j)=\varsigma_j^\top\xi_j-\|\varsigma_j\|^2/2$ exactly, and $|\varsigma_{j,i}|\le2\mub_i/\sigma_i$ gives
\[
\Lambda_t(\theta)\le\exp\Bigl(\sum_{j\le t:\,Z_j\ne\cem}\ \sum_i\frac{2\mub_i}{\sigma_i}|\xi_{j,i}|\Bigr).
\]
At a live step $\Sigma^{-1/2}(z_j-\mu_\theta(s_j))=\xi_j-\varsigma_j$, so \eqref{eq:score} and \Cref{ass:actor} give, for every $\theta$, $\|g_j\|\le\sum_i(B_{1,i}/\sigma_i)(|\xi_{j,i}|+2\mub_i/\sigma_i)$ and $\|\Upsilon_j\|\le\sum_i\bigl(B_{1,i}^2/\sigma_i^2+(B_{2,i}/\sigma_i)(|\xi_{j,i}|+2\mub_i/\sigma_i)\bigr)$. Since $\nabla_\theta\Lambda_t=\Lambda_tS_t$ and $\nabla^2_\theta\Lambda_t=\Lambda_t(S_tS_t^\top+\sum_{j\le t}\Upsilon_j)$, the norms of $\nabla^k_\theta\Lambda_t(\theta)$, $k=1,2$, are bounded for all $\theta$ by $\Psi_t:=\exp\bigl(\sum_{j\le t,i}(2\mub_i/\sigma_i)|\xi_{j,i}|\bigr)\,p_t(|\xi_0|,\dots,|\xi_t|)$, where $p_t$ is a polynomial with nonnegative coefficients determined by $t$ and the constants of \Cref{ass:actor}, and $\xi_j:=0$ at a dead step. Conditionally on $\mathcal F_j$, $\xi_j$ is standard Gaussian under $\theta_0$ at a live step, and a standard Gaussian $\xi$ has $\E[e^{c|\xi|}|\xi|^k]<\infty$ for all $c,k\ge0$; conditioning successively on $\mathcal F_t,\dots,\mathcal F_0$ therefore gives $\E_{\theta_0}\Psi_t<\infty$. Hence $|r_t|\,\|\nabla^k_\theta\Lambda_t(\theta)\|\le\Rmax\Psi_t$ for all $\theta$, and differentiation under the integral gives \eqref{eq:lr} at $\theta=\theta_0$ with continuous derivatives. At a live step $z_j\sim\mathcal N(\mu_\theta(s_j),\Sigma)$ given $\mathcal F_j$, so $\E[g_j\mid\mathcal F_j]=0$; the moment bounds are \Cref{prop:actor}; at dead steps $g_j=\Upsilon_j=0$. Since $g_i$ is $\mathcal F_j$-measurable for $i<j$, $\E[g_i^\top g_j]=0$ and $\E\|S_t\|^2=\sum_{j\le t}\E\|g_j\|^2\le(t+1)G^2$. Since $\E_{\theta_0}[\Lambda_t(\theta)]=1$ for every $\theta$, the same differentiation gives $\E_E[S_t]=0$ and $\E_E[S_tS_t^\top+\sum_{j\le t}\Upsilon_j]=0$, so $r_t$ may be replaced by $r_t-\Rmax/2$ in \eqref{eq:lr}; as $|r_t-\Rmax/2|\le\Rmax/2$ on every history, including absorbed ones, $\|\nabla\E_E[r_t]\|\le\tfrac12\Rmax G\sqrt{t+1}$ and $\|\nabla^2\E_E[r_t]\|\le\tfrac12\Rmax(t+1)(G^2+L_g)$ uniformly in $\theta$. Since $\sum_t\gamma^t(t+1)=(1-\gamma)^{-2}$, the derivative series converge uniformly on $\R^p$, and uniform convergence of derivatives~\cite[Thm.~7.17]{rudin1976}, applied to each partial derivative along coordinate segments, gives termwise first and second partial derivatives of $J_E$; these are uniform limits of continuous functions, so $J_E\in C^2$ with the bound $L_J$.
\end{proof}

\subsection{From model error to gradient error}\label{sec:transfer}
\begin{lemma}[Local law comparison]\label{lem:local}
Under \Cref{ass:env}, for every live $(Z,u)$ with $Z=(x,m)$, the kernels $\mathcal K_{\real}(\cdot\mid Z,u)$ and $\mathcal K_{\sm}(\cdot\mid Z,u)$ are mutually absolutely continuous and
\begin{multline}\label{eq:q}
\KL\bigl(\mathcal K_{\real}(\cdot\mid Z,u)\,\big\|\,\mathcal K_{\sm}(\cdot\mid Z,u)\bigr)\le \varrho(x,u)\\:=\tfrac12\|f_{\real}-f_{\sm}\|_{Q^{-1}}^2(x,u)\\+\tfrac12\E_{\real}\bigl[\mathbf 1\{x^+\in\Dx\}\|h_{\real}-h_{\sm}\|_{R_v^{-1}}^2(x^+)\bigm|x,u\bigr]\le\tfrac{\kappa^2}{2},
\end{multline}
where $x^+\sim\mathcal N(f_{\real}(x,u),Q)$. At $\cem$ the two kernels coincide.
\end{lemma}
\begin{proof}
Fix a live $(Z,u)$ and draw $X^+\sim\mathcal N(f_E(x,u),Q)$, then $Y^+\sim\mathcal N(h_E(X^+),R_v)$ if $X^+\in\Dx$ and $Y^+=\bot$ otherwise, for a fixed dummy symbol $\bot$. The common measurable map that sends $Y^+=\bot$ to $\cem$, and otherwise sends $(X^+,Y^+)$ to $(X^+,F(m,u,Y^+))$ when this pair belongs to $D$ and to $\cem$ otherwise, pushes the auxiliary law under $E$ forward to $\mathcal K_E(\cdot\mid Z,u)$. The physical Gaussian laws share the covariance $Q$, the observation Gaussian laws share $R_v$ whenever $X^+\in\Dx$, and both conditional observation laws put all their mass on $\bot$ when $X^+\notin\Dx$, so the auxiliary laws are mutually absolutely continuous, and so are their pushforwards. The chain rule and $\KL(\mathcal N(\mathsf a,W)\|\mathcal N(\mathsf b,W))=\tfrac12\|\mathsf a-\mathsf b\|_{W^{-1}}^2$ give $\KL_{\rm aux}=\varrho(x,u)$. The data-processing inequality~\cite[Thm.~2.17]{polyanskiy2025} yields \eqref{eq:q}. Since $x\in\Dx$ at a live state, \eqref{eq:eps} gives $\varrho\le\kappa^2/2$. At $\cem$ both kernels are the point mass at $\cem$.
\end{proof}
\needspace{5\baselineskip}
\begin{theorem}[Gradient transfer]\label{thm:transfer}
Under \Cref{ass:env,ass:actor}, for every $\theta\in\R^p$ and every $H\in\N\cup\{\infty\}$,
\begin{align}
	\begin{aligned}
		\bigl\|\nabla J_{\real}^H(\theta)-\nabla J_{\sm}^H(\theta)\bigr\|&\le\frac{\Rmax G\kappa}{\sqrt2}\sum_{t=1}^{H-1}\gamma^t\sqrt{t(t+1)}\\
		&\le\frac{\Rmax G\kappa}{\sqrt2\,(1-\gamma)^2}.
	\end{aligned}\label{eq:transfer}
\end{align}
\end{theorem}
\begin{proof}
Fix $t\ge1$ and let $\mathbb P_t,\mathbb Q_t$ be the laws of $\omega_t$ under $(\real,\theta)$ and $(\sm,\theta)$. They share the initial law and the policy kernels and differ only through the $t$ environmental kernels, which are mutually absolutely continuous by \Cref{lem:local}; hence $\mathbb P_t\ll\mathbb Q_t$ and the chain rule~\cite[Thm.~2.16(c)]{polyanskiy2025} gives
\begin{align}
	\begin{aligned}
		\KL(\mathbb P_t\|\mathbb Q_t)&=\sum_{j<t}\E_{\real}\bigl[\KL\bigl(\mathcal K_{\real}\|\mathcal K_{\sm}\bigr)(Z_j,u_j)\bigr]\\
		&\le\sum_{j<t}\bar\varrho_j\le\frac{t\kappa^2}{2},\\
		\bar\varrho_j&:=\E_{\real}[\mathbf 1\{j<\tau\}\varrho(x_j,u_j)].
	\end{aligned}\label{eq:KLt}
\end{align}
The inequality follows from \Cref{lem:local}, and absorbed transitions contribute zero because their kernels coincide. The quantity $\Gamma_t=(r_t-\Rmax/2)S_t$ is the same measurable function of $\omega_t$ under both laws, $\nabla\E_E[r_t]=\E_E\Gamma_t$ because $\E_E[S_t]=0$, and by \Cref{lem:score} $\E_{\mathbb P_t}\|\Gamma_t\|^2\le\tfrac14\Rmax^2(t+1)G^2$, with the same bound under $\mathbb Q_t$. For laws $\mathbb P,\mathbb Q$ with densities $\mathsf p,\mathsf q$ with respect to $\mathbb P+\mathbb Q$, let $\mathsf H^2(\mathbb P,\mathbb Q)=\int(\sqrt{\mathsf p}-\sqrt{\mathsf q})^2\in[0,2]$ be the squared Hellinger distance, and let $s_2=\E_{\mathbb P}\|\Gamma\|^2+\E_{\mathbb Q}\|\Gamma\|^2<\infty$. Write $\E_{\mathbb P}\Gamma-\E_{\mathbb Q}\Gamma=\int\Gamma(\sqrt{\mathsf p}-\sqrt{\mathsf q})(\sqrt{\mathsf p}+\sqrt{\mathsf q})$. The triangle inequality, Cauchy--Schwarz and $(\sqrt{\mathsf p}+\sqrt{\mathsf q})^2\le2(\mathsf p+\mathsf q)$ give $\|\E_{\mathbb P}\Gamma-\E_{\mathbb Q}\Gamma\|\le\sqrt{2s_2}\,\mathsf H(\mathbb P,\mathbb Q)$, and Jensen gives $\KL(\mathbb P\|\mathbb Q)\ge-2\log\int\sqrt{\mathsf p\mathsf q}=-2\log(1-\mathsf H^2/2)\ge\mathsf H^2$. Hence
\begin{align}
	\begin{aligned}
		\|\nabla\E_{\real}[r_t]-\nabla\E_{\sm}[r_t]\|&\le\Rmax G\sqrt{(t+1)\textstyle\sum_{j<t}\bar\varrho_j}\\
		&\le\frac{\Rmax G\kappa}{\sqrt2}\sqrt{t(t+1)} .
	\end{aligned}\label{eq:pert}
\end{align}
For $t=0$ the two expectations coincide. By \Cref{lem:score}, $\nabla J^H_E=\sum_{t<H}\gamma^t\nabla\E_E[r_t]$ for every $H$, which yields \eqref{eq:transfer}. Finally $\sqrt{t(t+1)}\le t+1$ and $\sum_{t\ge1}(t+1)\gamma^t=\gamma(2-\gamma)(1-\gamma)^{-2}\le(1-\gamma)^{-2}$.
\end{proof}
Note that no estimator-accuracy or contraction constant appears in \eqref{eq:transfer} because the observer enters only through the common pushforward in \Cref{lem:local}; the observer still shapes both objectives and the realized discrepancy through the closed-loop laws. Also note that the bound is worst-case and not uniform as the noise vanishes, since it uses the relative entropy of the full measurement history through $Q^{-1}$ and $R_v^{-1}$, and no matching lower bound is given.

\subsection{Occupancy and ascent consequences}\label{sec:cons}
We next replace the worst-case error in \eqref{eq:transfer} by errors under the plant's occupancy. For fixed $\theta$ define the occupancy-weighted dynamics and observation losses, which depend on $\theta$ through the plant law,
\begin{align}
\mathcal E_f&=(1-\gamma)\textstyle\sum_{j}\gamma^j\E_{\real}[\mathbf 1\{j<\tau\}\|f_{\real}-f_{\sm}\|_{Q^{-1}}^2(x_j,u_j)],\label{eq:Ef}\\
\mathcal E_h&=(1-\gamma)\textstyle\sum_{j}\gamma^j\E_{\real}\bigl[\mathbf 1\{j<\tau\}\mathbf 1\{x_{j+1}\in\Dx\}\nonumber\\&\qquad\qquad\qquad\times\|h_{\real}-h_{\sm}\|_{R_v^{-1}}^2(x_{j+1})\bigr],\label{eq:Eh}
\end{align}
with the observation term evaluated, as in \eqref{eq:q}, at the next physical state before the observer gate. By the tower property, $(1-\gamma)\sum_j\gamma^j \bar\varrho_j=\tfrac12(\mathcal E_f+\mathcal E_h)$ with $\bar\varrho_j$ as in \eqref{eq:KLt}, and $\mathcal E_f+\mathcal E_h\le\kappa^2$.

\begin{corollary}[Occupancy-weighted transfer]\label{cor:occ}
Under \Cref{ass:env,ass:actor}, with $a=1-\gamma$, for every $\theta$ and $H\in\N\cup\{\infty\}$,
\begin{multline}\label{eq:occ}
\bigl\|\nabla J_{\real}^H(\theta)-\nabla J_{\sm}^H(\theta)\bigr\|^2\\\le\min\Bigl\{\frac{\Rmax^2G^2\gamma(1-a^2)}{2a^4}\bigl(\mathcal E_f+\mathcal E_h\bigr)(\theta),\ \frac{\Rmax^2G^2}{a^3}\Bigr\}.
\end{multline}
\end{corollary}
\begin{proof}
Let $\bar\varrho_j=\E_{\real}[\mathbf 1\{j<\tau\}\varrho(x_j,u_j)]$ as in the proof of \Cref{thm:transfer}. Summing the first bound in \eqref{eq:pert} with discounting and applying Cauchy--Schwarz gives $(\sum_{t\ge1}\gamma^t\sqrt{(t+1)\sum_{j<t}\bar\varrho_j})^2\le(\sum_{t\ge1}\gamma^t(t+1))(\sum_{t\ge1}\gamma^t\sum_{j<t}\bar\varrho_j)$. The first factor is $a^{-2}-1=(1-a^2)/a^2$. By Tonelli, the second is $\sum_{j\ge0}\bar\varrho_j\sum_{t>j}\gamma^t=(\gamma/a)\sum_j\gamma^j\bar\varrho_j=\gamma(\mathcal E_f+\mathcal E_h)/(2a^2)$ by the tower identity. Multiplying by $\Rmax^2G^2$ gives the first term of \eqref{eq:occ}, and a finite $H$ only removes nonnegative terms. The centered form of \eqref{eq:lr} and Cauchy--Schwarz give $\|\nabla J_E^H\|\le\tfrac12\Rmax G\sum_t\gamma^t\sqrt{t+1}\le\tfrac12\Rmax Ga^{-3/2}$ for each environment, and $\|x-y\|^2\le2\|x\|^2+2\|y\|^2$ gives the second term.
\end{proof}
Note that the second term of \eqref{eq:occ} is a model-independent bound; the closed-form bound in \eqref{eq:transfer} improves on its square root only if $\kappa<\sqrt{2a}$, and the first term of \eqref{eq:occ} improves on it only if $\mathcal E_f+\mathcal E_h<2a/(\gamma(1-a^2))$.

\begin{corollary}[Ascent direction]\label{cor:ascent}
Under \Cref{ass:env,ass:actor}, let $d=\nabla J_{\sm}(\theta)$ and let $\delta\ge0$ satisfy $\|\nabla J_{\real}(\theta)-d\|\le\delta$. Valid choices of $\delta$ are the right-hand side of \eqref{eq:transfer} and the square root of that of \eqref{eq:occ}, with $H=\infty$. If $\|d\|>\delta$, then for every $0<\alpha<\tfrac{2}{L_J}\bigl(1-\delta/\|d\|\bigr)$,
\begin{multline}\label{eq:ascent}
J_{\real}(\theta+\alpha d)-J_{\real}(\theta)\\\ge\alpha\|d\|(\|d\|-\delta)-\tfrac12L_J\alpha^2\|d\|^2>0 .
\end{multline}
\end{corollary}
\begin{proof}
By \Cref{lem:score}, $J_{\real}$ is $L_J$-smooth, so $J_{\real}(\theta+\alpha d)-J_{\real}(\theta)\ge\alpha\langle\nabla J_{\real}(\theta),d\rangle-\tfrac12L_J\alpha^2\|d\|^2$, and $\langle\nabla J_{\real}(\theta),d\rangle=\|d\|^2+\langle\nabla J_{\real}(\theta)-d,d\rangle\ge\|d\|(\|d\|-\delta)$. The right side of \eqref{eq:ascent} is positive exactly when $\alpha<2(\|d\|-\delta)/(L_J\|d\|)$.
\end{proof}

\needspace{5\baselineskip}
\section{Finite-Budget Training Guarantee}\label{sec:train}
Policy optimization uses the causal estimator~\cite{baxter2001,yuan2022}
\begin{equation}\label{eq:est}
\widehat g_K=\sum_{t=0}^{K-1}\gamma^tr_tS_t .
\end{equation}
\emph{Algorithm 1.} Given the frozen $(f_{\sm},h_{\sm},F,D,\nu_0,r)$, integers $T,B,K\ge1$, $c>0$ and $\theta_0$, set $\alpha=c/\sqrt T$; for $t=0,\dots,T-1$ run $B$ independent stopped rollouts of length at most $K$ in $\sm$ under $\theta_t$ with fresh randomness and resets from $\nu_0$, form the batch average $\bar g_t$ of \eqref{eq:est}, and set $\theta_{t+1}=\theta_t+\alpha\bar g_t$; output $\theta_{\iota}$ with $\iota\sim\mathrm{Unif}\{0,\dots,T-1\}$ drawn independently~\cite{ghadimi2013}. No critic, advantage normalization, gradient clipping or data-dependent stopping of training is used.

\begin{lemma}[Estimator]\label{lem:est}
Under \Cref{ass:env,ass:actor}, in either environment $E$, $\E_E[\widehat g_K\mid\theta]=\nabla J_E^K(\theta)$,
\begin{align}
	\begin{aligned}
		\|\nabla J_E^K-\nabla J_E\|&\le b_K:=\tfrac12\Rmax G\sum_{t\ge K}\gamma^t\sqrt{t+1}\\
		&\le\bar b_K:=\frac{\Rmax G\gamma^K\sqrt{1+K(1-\gamma)}}{2(1-\gamma)^{3/2}},
	\end{aligned}\label{eq:bK}
\end{align}
and $\E_E[\|\widehat g_K\|^2\mid\theta]\le M_2:=\Rmax^2G^2/(1-\gamma)^3$, the second term of \eqref{eq:occ}; the batch average in Algorithm~1 satisfies the same identities and bounds.
\end{lemma}
\begin{proof}
Unbiasedness is \eqref{eq:lr} summed over $t<K$. The tail satisfies $\|\sum_{t\ge K}\gamma^t\nabla\E[r_t]\|\le\tfrac12\Rmax G\sum_{t\ge K}\gamma^t\sqrt{t+1}=b_K$ by \Cref{lem:score}, and Cauchy--Schwarz with $\sum_{t\ge K}\gamma^t=\gamma^K/(1-\gamma)$ and $\sum_{t\ge K}\gamma^t(t+1)=\gamma^K(1+K(1-\gamma))/(1-\gamma)^2$ gives $b_K\le\bar b_K$. By Minkowski and Cauchy--Schwarz, $(\E\|\widehat g_K\|^2)^{1/2}\le\Rmax G\sum_{t\ge0}\gamma^t\sqrt{t+1}\le\Rmax G(1-\gamma)^{-3/2}$. By Jensen, the batch average satisfies the same bounds; no reduction with $B$ is exploited.
\end{proof}

\begin{proposition}[Deployment stationarity]\label{thm:deploy}
Under \Cref{ass:env,ass:actor}, let $\theta_{\iota}$ be the output of Algorithm~1 with $\alpha=c/\sqrt T$ for any fixed $c>0$, and let $\Delta_J\ge\sup_\theta J_{\sm}(\theta)-J_{\sm}(\theta_0)$, for which $\Delta_J=\Rmax/(1-\gamma)$ is always admissible. With the expectation over the training rollouts and the output index,
\begin{equation}\label{eq:deploy}
\E\|\nabla J_{\real}(\theta_{\iota})\|^2\le\frac{4\Delta_J/c+2L_JcM_2}{\sqrt T}+2b_K^2+\frac{\Rmax^2G^2}{(1-\gamma)^4}\kappa^2 .
\end{equation}
The last term may be replaced by twice the expectation over training of the first term of \eqref{eq:occ} at $\theta_{\iota}$.
\end{proposition}
\begin{proof}
The argument is the randomized-output analysis of~\cite{ghadimi2013} with a biased estimator, as in~\cite{ajalloeian2020}. Write $J=J_{\sm}$ and $\bar g_t$ for the batch average at iteration $t$. \Cref{lem:est} gives $\E[\bar g_t\mid\theta_t]=\nabla J(\theta_t)+e_t$ with $\|e_t\|\le b_K$ and $\E[\|\bar g_t\|^2\mid\theta_t]\le M_2$. Since $\nabla J$ is $L_J$-Lipschitz by \Cref{lem:score}, the descent lemma gives $J(\theta_{t+1})\ge J(\theta_t)+\alpha\langle\nabla J(\theta_t),\bar g_t\rangle-\tfrac{L_J\alpha^2}{2}\|\bar g_t\|^2$. Taking conditional expectations and using $\langle\nabla J,e_t\rangle\ge-\tfrac12\|\nabla J\|^2-\tfrac12b_K^2$ gives $\E[J(\theta_{t+1})-J(\theta_t)\mid\theta_t]\ge\tfrac\alpha2\|\nabla J(\theta_t)\|^2-\tfrac\alpha2b_K^2-\tfrac{L_J\alpha^2}{2}M_2$. No restriction on $\alpha$ is needed because $M_2$ bounds the full second moment. Summing over $t<T$, taking expectations, using $\E J(\theta_T)-J(\theta_0)\le\Delta_J$ and dividing by $T\alpha/2$ gives $\E\|\nabla J(\theta_{\iota})\|^2=\frac1T\sum_{t<T}\E\|\nabla J(\theta_t)\|^2\le\frac{2\Delta_J}{\alpha T}+b_K^2+L_J\alpha M_2$, which with $\alpha=c/\sqrt T$ equals $(2\Delta_J/c+L_JcM_2)/\sqrt T+b_K^2$. Finally $\|\nabla J_{\real}\|^2\le2\|\nabla J_{\sm}\|^2+2\|\nabla J_{\real}-\nabla J_{\sm}\|^2$, and applying the squared bound in \Cref{thm:transfer} gives \eqref{eq:deploy}; applying \Cref{cor:occ} at $\theta_{\iota}$ instead gives the occupancy-weighted alternative.
\end{proof}

Note that \Cref{thm:deploy} does not establish monotone improvement, last-iterate convergence, optimality or closed-loop stability. For $M_2>0$ and a target $\eta>0$ (if $M_2=0$, all gradients vanish), since $\bar b_K^2\le\tfrac14M_2e^{-(1-\gamma)K}$, if the model term is at most $\eta/3$, then $K=\max\{1,\lceil(1-\gamma)^{-1}\log(3M_2/(2\eta))\rceil\}$ and $T\ge9(4\Delta_J/c+2L_JcM_2)^2/\eta^2$ give $\E\|\nabla J_{\real}(\theta_{\iota})\|^2\le\eta$ with at most $BTK$ simulator transitions and no plant interaction after identification; a fixed $\kappa>0$ limits the accuracy $\eta$ certified by \eqref{eq:deploy}, not the plant gradient itself.

\section{Numerical Evaluation}\label{sec:num}
\begin{figure*}[t]
\centering
\includegraphics[width=0.9\textwidth]{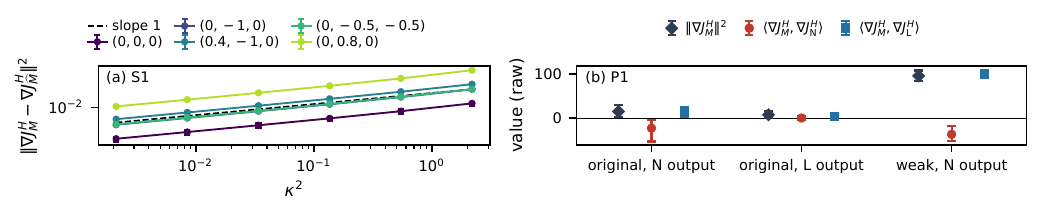}
\caption{Gradient comparisons, $H=512$, raw discounted objective. (a) S1 squared discrepancy against $\kappa^2=(1249/576)\lambda^2$ for five fixed policies under the nominal EKF (T1, T2); dashed line of slope one; 95\% $t$ intervals over 20 blocks. (b) P1 mean inner products between the plant gradient and the nominal (N) and learned (L) simulator gradients at common frozen parameters under $F_\ell$, the first-seed random outputs of the runs named on the axis (T6); plant squared gradient as reference; 95\% cluster-bootstrap intervals over twelve datasets.}\label{fig:e1}
\end{figure*}
\subsection{Shared protocol}\label{sec:setup}
To illustrate our results, we use the EKF as an example of the shared state estimator for a scalar system (S1) and a pendulum (P1). Both plants are simulated, with true states available for calibration and evaluation but never to the controller. Nominal and learned EKFs use nominal and identified model maps, respectively. Each plant--simulator pair shares a fixed EKF recursion, with estimates and covariances evolving separately. S1 provides T1--T5 at fixed policies, and P1 provides T6 and T7 through training (\Cref{tab:bench}). T1 checks the implementation, T2 and T5 compare estimated gradient discrepancies with bounds, and T4 evaluates bound magnitudes using plug-in diagnostics. T3, T6 and T7 report empirical observations. Code reproducing all reported results is available from the authors.

\emph{Systems.} S1 has $f_0(x,u)=0.7x+0.2u$, $h_0(x)=x$, $f_{\real}=f_0+0.08\sin x+0.02\tanh u$ and $h_{\real}=h_0+0.15+0.08\sin x$. P1 has state $(q,\omega)$ with $q=0$ upright and no angle wrapping, $q^+=q+0.05\,\omega+w_q$, $\omega^+=\omega+0.05\,(9.81\sin q-0.1\,\omega+u)+0.2\,(0.15\sin q-0.10\,\omega-0.05\tanh u)+w_\omega$, $y^+=q^++0.20+0.10\sin q^++v$ and $u\in(-6,6)$; the nominal model omits the residual dynamics term and the observation residual $0.20+0.10\sin q^+$.

\emph{Time scales, identification and tuning.} With sampling period $0.05$\,s and $\gamma=0.95$ the discount time constant is $0.97$\,s; survival is $\Pr(\tau>512)$ over a $25.6$\,s episode, and evaluation retains the stochastic training policy. Identification uses privileged reset data, $512$ transitions and $512$ observations per dataset plus $128$ validation samples of each, drawn uniformly from the action box and a state box $10\%$ wider than the physical gate, and Mat\'ern-$5/2$ kernel ridge regressions with length scale and ridge parameter selected by validation error, fitting on P1 the velocity residual of the dynamics and the residual of the observation map; these supervised fits do not optimize the occupancy-weighted loss of \Cref{cor:occ}, whose plug-in evaluation below is a diagnostic. The step coefficient $c$ of Algorithm~1 was selected from $\{0.03,0.1,0.3\}$ by simulator-only return and frozen at $0.03$.

\begin{table}[t]
\centering\scriptsize
\caption{Benchmarks. Both systems use $\gamma=0.95$, rewards $r\in(0,1]$ and $\Rmax=1$.}\label{tab:bench}
\renewcommand{\arraystretch}{0.9}\setlength{\tabcolsep}{3pt}\begin{tabular}{@{}p{0.8cm}>{\raggedright\arraybackslash}p{2.95cm}>{\raggedright\arraybackslash}p{4.45cm}@{}}\toprule
 & S1 & P1\\\midrule
noise & $Q=0.03^2$, $R_v=0.08^2$ & $Q=\diag(10^{-4},10^{-3})$, $R_v=10^{-2}$\\
live set & $|x|<1.5$, $|\xhat|<2$, $10^{-6}\le P\le4$ & $|q|<1.2$, $|\omega|<4$, $|\widehat q|<1.5$, $|\widehat\omega|<5$, $\operatorname{eig}(P)\subset[10^{-7},10]$\\
reset & $x_0\sim\mathrm U[-0.25,0.25]$, $\xhat_0=0$, $P_0=0.04$ & $q_0\sim\mathrm U[-0.1,0.1]$, $\omega_0\sim\mathrm U[-0.2,0.2]$, $\xhat_0=0$, $P_0=0.05I$\\
reward & $\exp(-2x^2-0.1u^2)$ & $\exp[-(q^2+0.1\omega^2+0.001u^2)]$\\
features & $(1,\tanh s,\tanh2s)/\sqrt3$ & \raggedright\arraybackslash $(1,\tanh(\widehat q/0.1),\tanh(\widehat\omega/0.5),$ $\tanh2\widehat q,\tanh(\widehat q\widehat\omega/2))/\sqrt5$\\
actor & $\mub=1.5$, $\sigma=0.35$, $u_{\max}=1$ & $\mub=1.5$, $\sigma=0.35$, $u_{\max}=6$\\
\bottomrule
\end{tabular}
\end{table}

\emph{Gradient measurement.} For each time $j$ a minimum-norm least-squares regression of the discounted reward-to-go on $[1,\phi(s_j)]$, fitted on an independent pilot batch ($4096$ trajectories on S1, $2048$ on P1) and frozen, is the predictable baseline $\beta(s_j,j)$. For horizon $H$ we measure $\widetilde g_H=\sum_{j<H}\gamma^jg_j[\sum_{t=j}^{H-1}\gamma^{t-j}r_t-\beta(s_j,j)]$, unbiased for $\nabla J^H$ since $\E[g_j\mid\mathcal F_j]=0$, with the same baseline on both sides of a pair. Conditional on the calibration data, the frozen parameter and the fitted baseline, in each of $n_{\rm b}$ independent blocks two independent halves of $n_{\rm p}$ trajectories or shared-noise trajectory pairs give $V_b=\bar v_1^\top\bar v_2$ with conditional mean $\|\E v\|^2$ exactly, where $v$ is a gradient estimate or a paired difference and $\E v$ its conditional mean; at random training outputs the average over datasets targets the expected squared quantity. The inner product of two gradients uses $\tfrac12(\bar g_1^\top\bar g'_2+\bar g_2^\top\bar g'_1)$. Negative estimates of nonnegative quantities are retained; the block standard error is that of $V_b$ over the blocks, and the S1 estimation error is the trajectory average of the discounted squared filter error $\sum_t\gamma^t\|\xhat_t-x_t\|^2$ before exit. Noise is shared within a pair, never between halves; pilot, measurement, training and evaluation streams are distinct, and training uses \eqref{eq:est} without checkpoint selection. A finite-difference check of the score estimator on S1 (perturbations \NumFDSteps) agrees to within \NumFDMaxZ\ standard errors. On P1 the calibration dataset is the independent unit. Paired effects are averaged over seeds within a dataset and summarized by a percentile cluster bootstrap over the twelve datasets ($10^4$ resamples), with a $t$ interval as sensitivity check; S1 cells use $t$ intervals over blocks. All intervals are approximate and pointwise.

\begin{figure*}[t]
\centering
\includegraphics[width=0.86\textwidth]{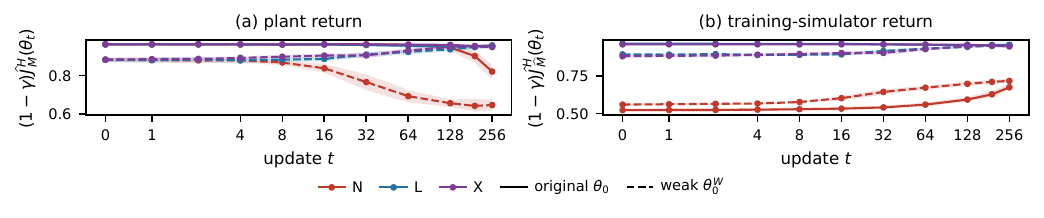}
\caption{P1 learning curves of frozen checkpoints, (a) plant and (b) training-simulator returns normalized by $1-\gamma$; solid and dashed curves are the original and weak initializations; means and 95\% cluster-bootstrap bands over twelve datasets; update axis linear through $2$, logarithmic thereafter.}\label{fig:learn}
\end{figure*}

\begin{table*}[t]
\centering\scriptsize
\caption{P1 outcomes at each method's own random output $\theta_{\iota}$, averaged within and across twelve datasets (all seeds); raw objective, $H=512$; intervals (95\% cluster bootstrap over datasets) only in the discrepancy column, which is zero by construction for X; RMS values pool live steps.}\label{tab:e2}
\setlength{\tabcolsep}{4pt}\begin{tabular}{@{}lccccc@{}}\toprule
policy & plant return & $\Pr(\tau>512)$ & RMS $q$ [rad] & RMS $u$ [N\,m] & $\|\nabla J_M^H-\nabla J_{\widehat M}^H\|^2$ \\
\midrule\multicolumn{6}{@{}l}{\emph{original initialization $\theta_0=(0,-2,-1,0,0)$}, 12 datasets, 3 seeds per dataset} \\
initial policy (reference) & $19.23$ & $0.56$ & $0.175$ & $3.54$ & -- \\
N, nominal simulator & $18.43$ & $0.27$ & $0.264$ & $3.41$ & $11$ $[-17,36]$ \\
L, learned simulator & $19.06$ & $0.48$ & $0.202$ & $3.48$ & $1$ $[0.2,2.1]$ \\
X, privileged exact simulator & $19.15$ & $0.45$ & $0.192$ & $3.47$ & $0$ (identity) \\
LQR, deterministic (reference) & $19.60$ & $0.98$ & $0.123$ & $2.57$ & n/a \\
LQR with exploration (reference) & $19.51$ & $0.97$ & $0.127$ & $3.13$ & n/a \\
\midrule\multicolumn{6}{@{}l}{\emph{weak initialization $\theta_0^W=(0,-0.7,-0.35,0,0)$}, 12 datasets, 2 seeds per dataset} \\
initial policy (reference) & $17.64$ & $0.00$ & $0.355$ & $2.78$ & -- \\
N, nominal simulator & $14.29$ & $0.00$ & $0.440$ & $2.54$ & $180$ $[140,220]$ \\
L, learned simulator & $18.77$ & $0.10$ & $0.279$ & $3.13$ & $5.1$ $[2.4,7.9]$ \\
X, privileged exact simulator & $18.86$ & $0.11$ & $0.265$ & $3.15$ & $0$ (identity) \\
\bottomrule\end{tabular}

\end{table*}

\subsection{Gradient discrepancy at fixed policies}\label{sec:e1}
We evaluate the five prespecified policies of \Cref{fig:e1}a under the frozen nominal EKF against $f_\lambda=f_{\real}+0.04\lambda\sin(\pi x/3)$ and $h_\lambda=h_{\real}+0.05\lambda\cos(\pi x/3)$ for $\lambda\in\{\NumEoneGrid\}$; on $\Dx=(-1.5,1.5)$ the suprema are $0.04\lambda$ and $0.05\lambda$, so $\kappa^2=(1249/576)\lambda^2$ exactly. We use $H=512$, for which $2\bar b_{512}<\NumTailFiveTwelve$, and $n_{\rm b}=20$ blocks with $n_{\rm p}=128$ pairs per half.

\emph{T1 and T2 (zero error and scaling).} At $\lambda=0$ the simulator is the plant, and under shared noise the paired gradients agree to floating-point precision although the estimation error is $0.11$ to $0.17$; this checks the pairing, not the theorem. For the 30 cells with $\lambda>0$ the estimated squared discrepancy is resolved and lies below the bound in every cell, and it increases in proportion to $\kappa^2$ with log--log slopes from \NumEoneSlopes. This scaling is compatible with a nonzero first-order response to the perturbation and does not establish sharpness.

\emph{T3 (observer and noise level).} At $\lambda=1$ we evaluate the policies $(0,-1,0)$ and $(0.4,-1,0)$ under three frozen observers. Their squared discrepancies are \NumObsNominal\ under the nominal EKF, \NumObsLearned\ under an EKF built from a learned model, and \NumObsDetuned\ under the nominal EKF detuned by a measurement covariance of $25R_v$, while the estimation error ranges from \NumObsErrRange\ and is smallest for the learned EKF in both cases. Scaling $Q$ and $R_v$ by $\vartheta\in\{\NumNoiseScales\}$ at $\lambda=1$ for policy $(0,-1,0)$, with the EKF rebuilt from the scaled covariances, changes $\kappa^2$ from \NumNoiseKappaRange\ and the squared uniform bound by a factor of \NumNoiseBoundFactor\ as $\vartheta$ decreases from $4$ to $1/16$, while the measured squared discrepancy remains within \NumNoiseZRange; these settings vary the noise and the EKF jointly and do not isolate a zero-noise limit.

\emph{T4 (bound magnitudes).} \Cref{cor:occ} bounds the squared difference between the exact gradients; we compare its expressions with estimates of this discrepancy. For policy $(0,-1,0)$ at $\lambda=1$ the estimate is \NumEoneMeasured\ (95\% $t$ half-width). The estimated occupancy losses are about \NumEoneWeightedPct\ of $\kappa^2$ and give a plug-in first term of \eqref{eq:occ} of \NumEoneOccupancy, smaller by a factor of \NumOccGain\ than the squared closed-form bound \NumEoneUniform. Both exceed the model-independent cap $M_2$ of \NumEoneEnvelope. The plug-in minimum is therefore the cap, a valid bound whatever the population losses; the active term of the population minimum is not certified. The squared closed-form bound falls below the cap only if $\kappa<\NumKappaTrivial$ ($\lambda<\NumLamCross$ on S1, the three smallest positive tested values). For the nominal P1 simulator the supremum over the open state and action boxes, approached at their corners, is $\kappa^2=\NumPoneKappaTot$, with squared bound $\NumPoneUniformN$. For the learned simulators the grid substitutes for $\kappa^2$ (maxima over an evaluation grid of the noise-weighted squared dynamics and observation errors, summed as in \eqref{eq:eps}) range from \NumPoneKappaLRange\ (mean \NumPoneKappaLTot) and are not certified suprema; their mean plug-in value $\NumPoneUniformL$ of the squared transfer expression is also above $M_2$. The model term of \eqref{eq:deploy} is below $M_2$ only when $\kappa^2<\NumDeployKappaTwo$, and the optimization term at the P1 budget is $\NumOptTermP$, while the truncation term is negligible, so the transfer and deployment expressions provide no useful P1 certificate.

\emph{T5 (truncation).} The proof of \Cref{lem:est} gives $\|\nabla J^K_{\real}-\nabla J^{512}_{\real}\|\le\tfrac12\Rmax G\sum_{t=K}^{511}\gamma^t\sqrt{t+1}\le b_K$. For \NumEfourPols\ S1 policies on the plant and $K\in\{\NumEfourKs\}$, the upper interval end of the measured squared difference lies below the square of this finite-horizon bound in all \NumEfourCells\ cells, by a factor of at least \NumEfourMinFactor.

\subsection{Training outcomes on the pendulum}\label{sec:e2}
On P1 we train in the nominal simulator N, the residual-augmented simulator L and the privileged exact-model simulator X under the frozen learned EKF $F_\ell$ of calibration dataset $\ell$, for twelve datasets, with three optimizer seeds per dataset and paired initialization $\theta_0+\mathcal N(0,10^{-4}I)$, $\theta_0=(0,-2,-1,0,0)$. An exploratory weak-initialization arm, added after observing that $\theta_0$ is already close to the control reference, uses $\theta_0^W=0.35\,\theta_0$ with the same perturbation, two seeds per dataset and fresh optimizer streams; training streams and output indices are independent across N, L and X. Every run of Algorithm~1 uses $T=256$, $B=16$, $K=256$, $\alpha=0.03/16$ and the uniform random output $\theta_{\iota}$; at eleven checkpoints the frozen iterate is evaluated on 128 held-out episodes of horizon $512$ in the plant and in its training simulator under shared evaluation noise. The initial policy and an observer-fed saturated linear-quadratic regulator (LQR) are references under the same $F_\ell$ and episodes; the LQR is designed in discrete time on the nominal linearization with the reward weights $\diag(1,0.1)$, $0.001$ and gain $K_{\rm LQR}=(\NumLQRGain)$, its latent mean is $\operatorname{artanh}(v)$ with $v$ the normalized torque $-K_{\rm LQR}\xhat/u_{\max}$ clipped to $[-\tanh1.5,\tanh1.5]$, and it is applied as $u=u_{\max}\tanh(\cdot)$ without and with the exploration noise $\sigma\xi$ added to the latent mean. Gradient measurements on P1 use $n_{\rm b}=8$ blocks of $n_{\rm p}=32$ pairs per half; \Cref{tab:e2} uses all seeds at each run's own output, and \Cref{fig:e1}b, T6 and T7 use the first seed at the random outputs of the N and L runs.

\emph{T6 (alignment at common parameters).} \Cref{cor:ascent} gives a sufficient ascent condition at a fixed parameter. We estimate the plant gradient $g$ and the nominal- and learned-simulator gradients $d$ at the same frozen parameters under $F_\ell$, the random outputs of the first optimizer seed of each dataset (\Cref{fig:e1}b), and evaluate the plug-in margin $\|d\|^2-\|g-d\|^2$ and the inner product $\langle g,d\rangle$ per dataset. At the nominal run's output in the weak-initialization arm the nominal-simulator margin is negative in \NumDMargNegNNRW\ of \NumDDatasets\ datasets (block $t$ interval below zero in \NumDMargNegResNNRW) and its inner product is negative in \NumDInnerNegNNRW\ (resolved in \NumDInnerNegResNNRW), with mean \NumDInnerNNRW; the learned-simulator margin is positive in \NumDMargPosLNRW\ of \NumDDatasets\ (resolved in \NumDMargPosResLNRW) and its inner product is positive in \NumDInnerPosLNRW\ (resolved in \NumDInnerPosResLNRW), with mean \NumDInnerLNRW. The original arm shows the same signs at the nominal run's output (inner products negative for N and positive for L in \NumDInnerNegNNR\ and \NumDInnerPosLNR\ of \NumDDatasets\ datasets; means \NumDInnerNNR\ and \NumDInnerLNR). A negative margin does not imply negative alignment; the inner products carry the evidence. The margins use estimated discrepancies, whereas the analytical nominal-model radius and the learned-model plug-in radii, the square roots of the corresponding squared expressions of T4, exceed every measured gradient norm, and $\alpha=0.03/16$ exceeds the ceiling $2/L_J=\NumAlphaCert$ of the sufficient step range, so \Cref{cor:ascent} does not certify the training updates.

\needspace{3\baselineskip}
\emph{T7 (plant gradient at the training output).} \Cref{thm:deploy} bounds the expected squared plant gradient at $\theta_{\iota}$ by optimization, truncation and model terms; the nominal and learned runs share $T$, $K$, $B$ and the admissible $\Delta_J=\Rmax/(1-\gamma)$, so only the model term differs, $\NumPoneModelTermN$ for N against the plug-in value $\NumPoneModelTermL$ for L, which is a diagnostic rather than a bound. The estimated squared plant gradients at the first-seed random outputs of T6 are \NumDNormMNR\ (N) against \NumDNormMLR\ (L) in the original arm and \NumDNormMNRW\ against \NumDNormMLRW\ in the weak-initialization arm, with paired N$-$L differences \NumDNormMDiffNR\ and \NumDNormMDiffNRW\ over datasets; an ordering of upper bounds does not imply the same ordering of the exact gradient norms. A small plant gradient measures parameter stationarity and does not by itself imply good control; the sensitivity $\operatorname{sech}^2(\theta^\top\phi(s))$ of the saturating actor tends to zero as $|\theta^\top\phi(s)|$ increases, so saturation can reduce parameter sensitivity, and return and survival are therefore reported separately.

\emph{Plant return.} In \Cref{fig:learn}, from the original initialization the initial plant return is \NumInitRetD, and its per-dataset paired change at $t=256$ is \NumChgND\ for N, \NumChgLD\ for L and \NumChgXD\ for X, while the nominal simulator's own return increases by \NumSimRiseND. From the weak initialization the initial plant return is \NumInitRetW, the changes are \NumChgNW, \NumChgLW\ and \NumChgXW, and the nominal simulator's return increases by \NumSimRiseNW. Even the privileged exact simulator shows no resolved improvement from the original initialization, so transfer error alone does not explain that outcome. The paired L$-$N plant-return effect at the random outputs (\Cref{tab:e2}) over twelve datasets is \NumEffRetD\ ($t$ interval \NumEffRetDt) for the original and \NumEffRetW\ (\NumEffRetWt) for the weak initialization. The LQR reference (\NumLQRDet) exceeds every training method (\Cref{tab:e2}) at this budget.

\emph{Observer dependence.} The sensor bias that $F_\ell$ was fitted to remove enters \Cref{thm:transfer} through the observation term ($\kappa_h^2=\NumPoneKappaH$ of $\kappa^2=\NumPoneKappaTot$), not through the estimator. With the nominal EKF shared between N and the plant instead, the RMS angle-estimation error of the initial policy over live evaluation steps is \NumNomRMSEqD\ rad against \NumRMSEqInitD\ under $F_\ell$, every episode of the initial policy and of the LQR reference exits through the physical gate (returns \NumNomInitRetD\ and \NumNomLQRD), and training in N changes the plant return at $t=256$ by \NumNomChgND\ against \NumNomChgXD\ for X ($t$ intervals over \NumNomRunsD\ runs); the return-change pattern thus depends on the shared estimator, which the uniform bound does not resolve.

\section{Conclusion and Future Work}\label{sec:concl}
We have studied policy-gradient transfer from a simulator to a plant when both are controlled through a shared fixed observer. We have shown that the difference between the plant and simulator policy gradients is bounded linearly by the noise-weighted dynamics and observation-model errors without an estimator-accuracy guarantee, and we have derived an occupancy-weighted bound, a sufficient ascent condition and a finite-budget bound on the expected squared plant gradient at a random iterate of training in the simulator alone. Simulation results on a scalar system and a pendulum have illustrated the bounds and their conservatism. On the pendulum, an identified residual model changes the estimated mean gradient alignment from negative to positive, while the bounds exceed the measured discrepancies by several orders of magnitude and do not certify the pendulum outcomes, and the realized discrepancy depends on the observer and the noise level, which the uniform bound does not resolve. Future work will consider sharper constants, finite-data certification of the model errors, unequal noise covariances, and quantitative links to control performance.

\bibliographystyle{IEEEtran}
\bibliography{refs}

\end{document}